\documentclass[aps,prl,amsmath,twocolumn,superscriptaddress]{revtex4}
\usepackage{bm}
\usepackage{graphicx}

\DeclareMathOperator{\Img}{\mathrm{Im}}
\DeclareMathOperator{\Sp}{\mathrm{Sp}}

\usepackage{cancel}

\begin{document}
\title{Triplet superconductivity in a ferromagnetic vortex}

\author{Mikhail S. Kalenkov}
\affiliation{I.E. Tamm Department of Theoretical Physics, P.N. Lebedev Physical Institute, 119991 Moscow, Russia}
\author{Andrei D. Zaikin}
\affiliation{Institute of Nanotechnology, Karlsruhe Institute of Technology (KIT), 76021 Karlsruhe, Germany}
\affiliation{I.E. Tamm Department of Theoretical Physics, P.N. Lebedev Physical Institute, 119991 Moscow, Russia}
\author{Victor T. Petrashov}
\affiliation{Department of Physics, Royal Holloway, University of London, Egham, Surrey TW20 0EX, United Kingdom}


\begin{abstract}
We argue that odd-frequency triplet superconductivity can be conveniently realized in
hybrid superconductor-ferromagnet (SF) structures with a ferromagnetic vortex. We demonstrate that due to proximity-induced long-range triplet pairing such SFS junctions can sustain appreciable supercurrent which can be directly measured in experiments.
\end{abstract}



\maketitle

A normal metal (N) sandwiched between two superconductors (S) can become superconducting as a result of penetration of Cooper pairs from the superconducting electrodes. The range of penetration is set by the so-called thermal length $\xi_T$ which can easily reach several micrometers at sufficiently low temperatures \cite{LR,BWBSZ,GKI}. The situation changes drastically if the normal metal is replaced by a ferromagnet (F). The quantum mechanical exchange interaction on the F-side then destroys conventional spin-singlet Cooper pairs within a few nanometers (the so-called paramagnetic effect) \cite{Panyukov}. Experiments to determine actual superconducting penetration depths in ferromagnets intensified more than a decade ago, when techniques were developed to fabricate hybrid nanoscale SF structures with well controlled geometries. Several groups \cite{Petrashov,Giroud,Petrashov2,Venkat,Petrashov3} reported an unexpectedly strong influence of superconductors that stimulated new theoretical efforts in a search for a sustainable superconductivity that is compatible with the exchange interaction. During the last decade several theoretical mechanisms were suggested \cite{BVE,KSJ,Eschrig03,11,Fominov07,Sasha,GKZ08}, some of which were successfully realized experimentally \cite{Petrashov1,Keizer,Aarts10,Blamire,Birge,Wang}. A recent comprehensive review of the status of the field was given in Ref. \cite{Eschrig11}.

Common to all mechanisms of long range proximity effect in ferromagnets is the generation of triplet superconductivity within highly inhomogeneous ferromagnetic regions adjacent to superconductors. The systems studied up to date include intrinsically inhomogeneous ferromagnets \cite{Petrashov1}, half-metallic ferromagnets with spin-active FS interfaces \cite{Keizer,Aarts10}, and engineered multilayers consisting of magnetic and non-magnetic materials \cite{Blamire,Birge}.

In this letter we address a different situation of proximity-induced long range triplet pairing in ferromagnets with magnetic vortex structure. Magnetic vortices are stable in systems intermediate between very small, 10 nm scale magnets, which behave as single giant spins, and macroscopic magnets with dimensions exceeding $\sim 1$ $\mu$m. The magnetic structure in such mesoscopic magnets is the result of a competition between exchange, anisotropy, and dipolar energies and depends strongly on their shape. The latter property allows magnetic nano-engineering using modern nanolithography \cite{Mironov10} opening possibility of investigating SFS structures with differing magnetic structures. Recently mesoscopic magnetic structures have attracted a lot of attention due to their remarkable transport properties \cite{Bruno04,Fraerman08,Neubauer09}, as well as the prospect of technological applications for magnetic storage of information of unprecedented density \cite{Cowburn99}. Below we will demonstrate that mesoscopic ferromagnetic structures can turn superconducting if attached to superconducting electrodes.

\begin{figure}
\centerline{
\includegraphics[width=80mm]{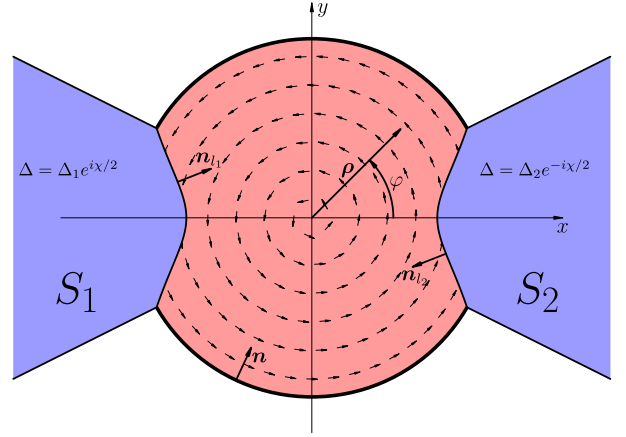}
}\caption{(Color online) SFS junction formed by two superconducting electrodes connected via ferromagnetic vortex.}
\label{sfvs-fig}
\end{figure}

{\it The model and quasiclassical formalism.} We will consider a ferromagnetic film of thickness $d$ located in the $xy$~plane with magnetization forming a vortex. This film is partially covered by two superconducting electrodes thus forming an SFS contact as it is shown in Fig. \ref{sfvs-fig}.  Our main goal is to analyze superconducting correlations that penetrate into a ferromagnetic vortex from
the electrodes. In order to accomplish this goal we will employ the quasiclassical Usadel equations \cite{Usadel,BWBSZ} for energy-integrated matrix Matsubara-Green functions $\check G$. E.g. in the ferromagnet with diffusion constant $D$ these equations read
\begin{equation}
iD\nabla (\check G \nabla \check G)=
[\check \Omega , \check G], \quad \check G^2=1,
\label{Usadel}
\end{equation}
where
\begin{gather}
\check G=
\begin{pmatrix}
\hat G & \hat F \\
\hat F^+ & \hat G^+ \\
\end{pmatrix}, \
\check \Omega=
\begin{pmatrix}
i\omega_n \hat 1 - \hat{\bm{\sigma}}\bm{h} & \Delta \hat 1\\
-\Delta^* \hat 1& - i\omega_n \hat 1 + \hat{\bm{\sigma}}\bm{h}\\
\end{pmatrix}
\label{Omega}
\end{gather}
are $4\times4$ matrices in Nambu and spin spaces. Their commutator in Eq. (\ref{Usadel}) and below is denoted by square brackets. Accordingly, $\hat G$, $\hat F$, $\hat F^+$ and $\hat G^+$ are $2\times 2$ matrices in the spin space, $\omega_n=\pi T(2n+1)$ is the Matsubara frequency, $\bm{h}$ is the exchange field in the ferromagnet and $\hat{ \bm{\sigma}}=(\hat\sigma_1,\hat\sigma_2,\hat\sigma_3)$ represents the Pauli matrices in the spin space. The same equations (\ref{Usadel}), (\ref{Omega}) hold also for superconducting electrodes, one should only replace $D$ by the diffusion constant in the corresponding electrode. The superconducting order parameter $\Delta$ equals to zero in the ferromagnet, while in two superconducting terminals it is respectively $\Delta =\Delta_1\exp (i\chi /2)$ and $\Delta =\Delta_2\exp (-i\chi /2)$ with real $\Delta_{1,2}$ and $\chi$ being the superconducting phase difference across our SFS junction.

Equations (\ref{Usadel}) should be supplemented by appropriate boundary conditions at each of the two SF-interfaces which account for electron transfer across these interfaces. In what follows we will assume that there exist tunnel barriers at both SF interfaces with the corresponding tunneling resistances $r_{1,2}$. In the tunneling limit it suffices to employ Kuprianov-Lukichev boundary conditions \cite{KL} at each SF-interface. E.g., at the interface
between the first superconducting electrode ($z>0$) and the ferromagnet ($z<0$) these boundary conditions read
\begin{equation}
2 r_1 \sigma  \check G_F \partial_z \check G_F=
[\check G_F, \check G_{S_1}],
\label{KupLuk}
\end{equation}
where $\hat G_F$ and $\hat G_{S_1}$ are respectively the Green functions at the F- and S-sides of the first interface and $\sigma$ is the Drude conductivity of a ferromagnet. Analogous boundary conditions hold for the second SF-interface.

{\it Long-range triplet pairing in a ferromagnetic vortex.} The presence of tunnel barriers at both SF-interfaces effectively implies weak electron tunneling regime in which case the proximity effect remains small and it suffices to linearize Usadel equations in the ferromagnet as
\begin{equation}
D\nabla^2\hat F -2\omega_n \hat F - i \{ \hat F , \bm{h}(\bm{r})\hat{\bm{\sigma}}\}=0.
\label{Feq}
\end{equation}
In Eq. (\ref{Feq}) we restrict Matsubara frequencies to be positive $\omega_n>0$ and denoted the anticommutator by curly brackets. A similar equation holds for the function $\hat F^+$.

In general magnetization patterns in thin ferromagnetic films depend on the film geometry and are influenced by the following trade-off. On one hand, magnetostatic energy minimum is reached provided the film magnetization remains in-plane. On the other hand, in some regions, such as, e.g., vortex cores, local magnetization can go out-of-plane in order to minimize the exchange energy. As the magnetic core radius typically remains small as compared to the superconducting coherence length, in the following we will assume that magnetization lies in-plane everywhere in the ferromagnet, see Fig. \ref{sfvs-fig}. In sufficiently thin films the exchange field $\bm{h}$ depends only on in-plane coordinates $(x,y)$ and can be represented as
$\bm{h}=(h \cos\theta, h \sin\theta)$
where $\theta =\theta (x,y)$. In this case the spin structure of the anomalous Green function $\hat F$ inside the ferromagnet can be chosen in the following form
\begin{equation}
\hat F= F_0+ \hat{\bm{\sigma}}\bm{m} F_h +  \hat{\bm{\sigma}} [\bm{e}_z,\bm{m}] F_t,
\label{comp}
\end{equation}
where $F_0$ describes the singlet pairing component, while $F_h$ and $F_t$ correspond to two different triplet components. In Eq.  (\ref{comp})
we also introduced in-plane and normal to the plane unity vectors $\bm{m}=\bm{h}/h$ and $\bm{e}_z$. Combining Eqs. \eqref{comp} and \eqref{Feq} we arrive at
the following equations for the above components:
\begin{gather}
D \nabla^2 F_0 -2 \omega_n F_0 = 2i h F_h,
\label{F0eq}
\\
\mathcal{D} F_h = D F_t \nabla_{\bm{\rho}} ^2\theta + 2 D (\nabla_{\bm{\rho}} F_t,\nabla_{\bm{\rho}} \theta) + 2i h F_0,
\label{Fheq}
\\
\mathcal{D}F_t =- D F_h \nabla_{\bm{\rho}}^2\theta -
2 D(\nabla_{\bm{\rho}} F_h,\nabla_{\bm{\rho}}\theta),
\label{Fteq}
\end{gather}
where we defined the differential operator
\begin{equation}
\mathcal{D}= D\nabla^2-D(\nabla_{\bm{\rho}} \theta )^2 -2\omega_n
\label{calD}
\end{equation}
and distinguished $\nabla$ and $\nabla_{\bm{\rho}}$ as respectively 3d and 2d (in-plane) gradient operators.

Note that Eqs.~\eqref{F0eq} and \eqref{Fheq} contain the exchange field $h$ thus providing the characteristic length scale both for $F_0$ and $F_h$ of order $\xi_h\sim \sqrt{D/h}$. At the same time, Eq.~\eqref{Fteq} does not contain the $h$-term and, hence, typical variations of $F_t$ occur on a much longer length scale $\xi_T \sim \sqrt{D/T} \gg \xi_h$. This observation illustrates the difference between the two triplet components $F_h$ and $F_t$ and constitutes the essence of the long range proximity effect in SFS structures: while
the components $F_0$ and $F_h$ decay already in the vicinity of an SF-interface, the
triplet component $F_t$ survives deep inside the ferromagnet provided the temperature remains sufficiently low.

Before turning to the solution of Eqs. \eqref{F0eq}-\eqref{Fteq} let us perform some further simplifications. Firstly, we will neglect both magnetic anisotropy and stray field effects. In this case outside the magnetic vortex core the function $\theta$ obeys the equation
\begin{equation}
\nabla_{\bm{\rho}}^2 \theta =0,
\label{thetaeq}
\end{equation}
which allows to drop the first terms in the right-hand side of Eqs. \eqref{Fheq} and \eqref{Fteq}. Secondly, we will assume the ferromagnetic film to be sufficiently thin $d \lesssim \xi_T$, in which case the dependence of the long-range triplet component $F_t$ on the coordinate $z$ can be neglected.
Then, integrating Eq. \eqref{Fteq} over $z$ we obtain
\begin{equation}
\mathcal{D}_{\bm{\rho}}F_t = -
2 D(\nabla_{\bm{\rho}} \overline{F}_h,\nabla_{\bm{\rho}}\theta),
\quad
\overline{F}_h = \dfrac{1}{d}\int_{-d}^0 F_h dz,
\label{Fteqav}
\end{equation}
where $\overline{F}_h$ is the average value of $F_h$ component over the ferromagnetic film thickness and  $\mathcal{D}_{\bm{\rho}}$ is defined by Eq. (\ref{calD}) with $\nabla^2\to \nabla_{\bm{\rho}}^2$.

Eq. (\ref{Fteqav}) accounts for diffusion of the long-range triplet component $F_t$ across the ferromagnet with nonuniform in-plane magnetization. It demonstrates that non-zero $F_t$ is generated in the parts of the ferromagnet where both $\nabla_{\bm{\rho}}\theta$ and $\nabla_{\bm{\rho}} \overline{F}_h$ differ from zero. The condition $\nabla_{\bm{\rho}}\theta \neq 0$ obviously holds everywhere in the ferromagnetic plane since the magnetization remains non-uniform there. As for the averaged component $\overline{F}_h$, it vanishes together with its gradient at distances exceeding $\sim \xi_h$ from SF-interfaces. In the immediate vicinity of such interfaces $\overline{F}_h$ is non-zero, but its in-plane gradient remains small because in the main approximation it only depends on the absolute value of the exchange field $h$, cf. Eq.~\eqref{Fheq}.  The gradient $\nabla_{\bm{\rho}} \overline{F}_h$ becomes appreciable only in the region of the ferromagnet just below the edge of the superconducting film where $\overline{F}_h$ changes abruptly. With this in mind we arrive at the following result for the long-range triplet component
\begin{equation}
F_t(\bm{\rho})=
\dfrac{iD^2}{h\sigma d}
\sum_{k=1,2} \dfrac{F_{S_k}}{r_k}
\int_{l_k} P^{\bm{\rho},\bm{\rho}'}_{\omega_n}
(\nabla^{\prime}_{\bm{\rho}}\theta(\bm{\rho}'), \bm{n}_{l_k}(\bm{\rho}'))d l_k,
\label{Ftres}
\end{equation}
which holds inside the ferromagnetic film. Here $F_{S_k}$ is anomalous Green function in the bulk of the $k$-th superconductor and $\bm{n}_{l_k}$ is the outer unity vector normal to the superconducting plane $S_k$ (see Fig.~\ref{sfvs-fig}). Integration contours $l_k$ in Eq.~\eqref{Ftres} are lines in the $xy$ plane corresponding to the edge of the superconductor $S_k$ and in the ferromagnet kernel $P^{\bm{\rho},\bm{\rho}'}_{\omega_n}$ obeys the equation
\begin{equation}
\mathcal{D}_{\bm{\rho}}
P^{\bm{\rho},\bm{\rho}'}_{\omega_n}= \delta(\bm{\rho}-\bm{\rho}'),
\end{equation}
with boundary conditions $\partial P^{\bm{\rho},\bm{\rho}'}_{\omega_n}/ \partial \bm{n} = 0$. We also note that Eq.~\eqref{Ftres} can easily be generalized to the case of arbitrary $\nabla_{\bm{\rho}}^2 \theta$ not obeying Eq.~\eqref{thetaeq}.

{\it Triplet pairing and Josephson effect.} As triplet pairing amplitude can survive deep in the ferromagnet, at sufficiently low temperatures our SFS junction can sustain appreciable supercurrent which is converted from singlet to triplet and back in the vicinity of SF-interfaces. In order to evaluate this supercurrent we will employ the standard expression for the current density
\begin{equation}
\bm{j}= \frac{\pi \sigma T}{2e}
\Img \sum_{\omega_n>0} \Sp [ \hat F\nabla\hat F^+ - \hat F^+ \nabla \hat F],
\label{current}
\end{equation}
where the trace is taken over the spin degree of freedom. Combining Eqs. \eqref{comp}, \eqref{Ftres} with \eqref{current} we recover the sinusoidal current-phase relation $I(\chi )=I_c\sin\chi$ with
\begin{gather}
I_c=\dfrac{2 \pi T D^3 }{ e h^2 \sigma d r_1 r_2}
\sum_{\omega_n>0}
\dfrac{\Delta_1\Delta_2}{\sqrt{(\omega_n^2+\Delta_{1}^2)(\omega_n^2+\Delta_{2}^2)}}
\label{IC}
\\
\times\int\limits_{l_1,l_2}
P^{\bm{\rho}_1,\bm{\rho}_2}_{\omega_n}
(\nabla_{\bm{\rho}}\theta(\bm{\rho}_1), \bm{n}_{l_1}(\bm{\rho}_1))
(\nabla_{\bm{\rho}}\theta(\bm{\rho}_2), \bm{n}_{l_2}(\bm{\rho}_2))
d l_1d l_2
\nonumber
\end{gather}
Note that in the course of our derivation we always assumed the proximity effect to be sufficiently weak. This assumption is satisfied under the condition
\begin{equation}
\dfrac{1}{r_{1,2}\sigma}\sqrt{\dfrac{D}{h}}\ll
\begin{cases}
1, \quad  & d  \gtrsim \xi_h,
\\
d\sqrt{h/D} , \quad  & d \lesssim \xi_h.
\end{cases}
\label{vc}
\end{equation}
Eq. (\ref{IC}) together with its validity condition (\ref{vc}) represents the central result of our analysis which fully determines the Josephson critical current of an SFS junction with a ferromagnetic vortex. Actually this result applies not only to vortex configurations but also to a broader class of non-uniform magnetization patterns.

Let us now assume that our ferromagnetic film has the form of a disk with radius $R$ and vortex-like magnetization pattern with vortex core located in the disc center. Then the function $\theta$ equals to $\varphi + \pi/2$ for clockwise or $\varphi - \pi/2$ for counterclockwise magnetization, where $\varphi$ is the azimuthal angle (see Fig. \ref{sfvs-fig}). Eq. \eqref{thetaeq} is fulfilled in this case. Remarkably, Eq. \eqref{IC} yields exactly the same result for quite different magnetization patterns: vortex-like ($\theta=\varphi \pm \pi/2$), antivortex-like ($\theta=-\varphi$ or $\theta=-\varphi + \pi$) and hedgehog-like ($\theta=\varphi$ or $\theta=\varphi + \pi$) states. This property holds since the function $\nabla_{\bm{\rho}}\theta$ remains the same (up to a sign) for all these magnetization patterns. Note, however, that for the last two patterns stray magnetic field is not confined to the disc center and may influence superconductivity in the electrodes.

\begin{figure}
\centerline{
\includegraphics[width=80mm]{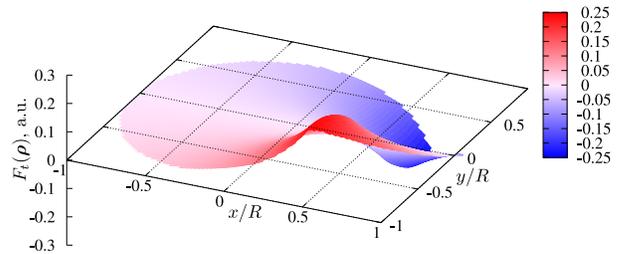}
}\caption{(Color online) Typical spatial distribution of the long-range superconducting triplet component $F_t$ induced in the ferromagnetic disk with vortex-like magnetization by one superconducting electrode ($x>R/2$, $z>0$) with real $\Delta$.}
\label{triplet-fig}
\end{figure}

For illustration, typical spatial profile of the long-range superconducting triplet component $F_t$ induced by one superconducting electrode in the ferromagnet with a vortex is schematically depicted in Fig. \ref{triplet-fig}. As it was expected, $F_t$ is most efficiently generated close to the edge of a superconductor where the scalar product $|(\nabla_{\bm{\rho}} \theta (\bm{\rho}), \bm{n}_{l} (\bm{\rho}))|$ reaches its maximum values. Provided the proximity effect remains weak, the total value of $F_t$ is given by a superposition of independent contributions from two superconducting electrodes, cf. also Eq. (\ref{Ftres}).

As one can observe in Fig. \ref{triplet-fig}, the long-range triplet component $F_t$ penetrating into the ferromagnet can take both positive and negative values. Thus,
depending on the magnetization pattern inside the ferromagnetic film it is possible to realize both zero- and $\pi$-junction states in our structure. The latter regime can be reached, e.g., by implementing certain asymmetry in SF contacts.

\begin{figure}
\centerline{
\includegraphics[width=80mm]{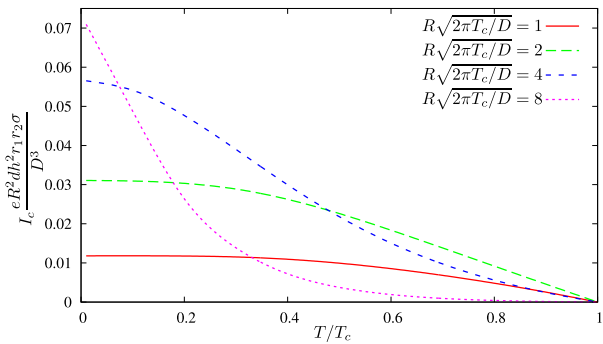}
}\caption{(Color online) $I_c$ versus temperature (normalized by the critical temperature $T_c$) in SFS junctions containing a ferromagnetic vortex at different values of $R$.
The edges of superconducting electrodes (contours $l_{1,2}$) are chosen to coincide with straight lines $y=\pm R/2$.
}
\label{J-T-fig}
\end{figure}

We further consider a symmetric situation, set $\Delta_{1,2}=|\Delta |$ and assume that the relevant Thouless energy $\varepsilon_{\mathrm{Th}} \sim D/(2R)^2$ remains smaller than the superconducting gap $|\Delta |$. Then in the limit $T \ll \varepsilon_{\mathrm{Th}}$ from Eq. \eqref{IC} we find
\begin{equation}
I_c \sim
D^2\varepsilon_{\mathrm{Th}}/(e d h^2 r_1 r_2 \sigma ),
\label{ic1}
\end{equation}
while at intermediate temperatures $\varepsilon_{\mathrm{Th}} \ll T \ll |\Delta|$ the  Josephson current follows the standard exponential dependence on temperature
\begin{equation}
I_c \sim  \dfrac{TD^2}{e d h^2 r_1 r_2 \sigma}
\exp\left(- L \sqrt{2\pi T/D}\right)
\label{ic2}
\end{equation}
where $L$ is an effective distance between the two SF contacts which depends on geometry details (obviously $L=2R$ for small area contacts). For illustration the Josephson critical current $I_c$ is also plotted in Fig. \ref{J-T-fig} as a function of temperature for different values of $R$.

Our result for $I_c$ in SFS systems turns out to be by the factor $\sim \varepsilon^2_{\mathrm{Th}}/h^2$ smaller than that for conventional diffusive SNS junctions with identical geometry, cf., e.g., \cite{GKI,KL}. The critical current of our SFS structure can further be increased by a proper choice of the system parameters. For a simple estimate of possible maximum values of $I_c$ let us employ Eq. (\ref{ic1}) at the border of its applicability range
(\ref{vc}). Then for $T \ll \varepsilon_{\rm Th}$ and $d \gtrsim \xi_h$ we obtain
\begin{equation}
I_c \sim
D^2\sigma/(eR^2dh) \sim (\xi_h/d)^2 \varepsilon_{\rm Th}/(eR_N),
\label{est}
\end{equation}
where $R_N$ is the normal state resistance of the ferromagnetic film between two superconducting electrodes. This estimate is also supported by our independent calculation (not presented here) which yields contributions to $I_c \propto 1/h$ in higher orders in barrier transmissions. Eq. (\ref{est}) demonstrates that for $d \gtrsim \xi_h$ one can expect to reach values of $I_c$ only by the factor $\sim \xi_h^2/d^2$ smaller that the absolute maximum
$I_c \sim \varepsilon_{\rm Th}/eR_N$ achieved for SNS junctions \cite{SNS}. Actually, the latter maximum value can also be reached, but only for extremely thin films $d \lesssim \xi_h$ (cf. Eqs. (\ref{ic1}), (\ref{vc})) with large values of $R_N$.

In summary, we demonstrated that long-range triplet superconductivity can coexist with a ferromagnetic vortex and evaluated the supercurrent across SFS junctions containing such vortex. For properly chosen system parameters the effect is well in the measurable range
and can be directly tested in future experiments. This work was supported in part by DFG, by RFBR under grant 09-02-00886 and by British EPSRC grant EP/F01689/1.

\end{document}